\documentclass[epsfig,12pt]{article}
\usepackage{epsfig}
\def\beqn{\begin{eqnarray}}
\def\eeqn{\end{eqnarray}}

\def\beq{\begin{equation}}
\def\eeq{\end{equation}}
\def\ba{\beq\new\begin{array}{c}}
\def\ea{\end{array}\eeq}

\begin{document}

\begin{flushright}
{\small FTPI-MINN-12/33, UMN-TH-3122/12}
\end{flushright}

\begin{center}
{  \large \bf  
FRONTIERS BEYOND THE STANDARD MODEL: \\[2mm]
Reflections and Impressionistic Portrait of the Conference\footnote{ 
 Frontiers Beyond the Standard Model,  FTPI,
 October 11-13, 2012.}.}
\end{center}

\begin{flushright}{\small Paraphrasing Feynman: Nature is more \\imaginative than any of us and all of us \\taken together. Thank god, it keeps \,\,\,\, \rule{0mm}{0mm}\\sending   messages rich on  surprises.
\quad\,\,\,\,\rule{0mm}{0mm}}
\end{flushright}

\vspace{0.6cm}

\begin{center}
{\large  
    M. Shifman}
   \end {center}

\begin{center}
{\it  
William I. Fine Theoretical Physics Institute, University of Minnesota,\\
Minneapolis, MN 55455, USA\\
shifman@umn.edu}

\end {center}



\section{In the beginning: before 1972}

At the beginning of my career in high energy physics (HEP), theory was lagging behind 
experiment and, by and large,  its development was guided by experiment. Before the advent 
of the standard model (SM, at that time referred to as the Weinberg-Salam model) 
and quantum chromodynamics (QCD), none of the hot theoretical topics of the 
day were particularly singled out: many directions of theoretical thought were considered 
to be equally respectable, and  peacefully coexisted. The HEP theory community was distributed roughly evenly 
between them. People understood that above several GeV theory had to 
be changed.\footnote{One of the reasons behind this understanding was the analysis 
of the $K_L$-$K_S$ mass difference which led to the GIM mechanism implying in turn the 
existence of a charmed quark not heavier than $\sim 2$ GeV.} A few competing ideas 
as to possible changes were discussed, but none were firmly established.
The ignorance at short distances was usually parametrized either by nonrenormalizable 
operators  in effective Lagrangians or, in loops, by an ultraviolet cut-off. Then 
the theoretical predictions were confronted with experiment in order to determine the scale 
of ``new physics." Experiment was an ultimate judge of what was important in theory and what not; 
I do not think that anybody could even dream of making a statement that ``the theory 
of everything" was within reach. The general belief was that such a theory (that would explain all 
mysteries of nature once and for all) could not exist. For this reason a lack of advancement, 
or even complete failure of a given line of thought, caused no trouble in the 
community. It only affected a few followers who could rather painlessly switch to 
other theories or topics. This was a wonderful time.

Experimental guidance started fading  away after the November revolution of 1974 -- the 
discovery of heavy charmonium. The role played by experiment continued to decrease  
steadily for quite some time, until it became almost invisible. In HEP theory this effect 
coincided with a transition (a crossover, rather than a phase transition) into a different 
mode of operation which I will call, somewhat conditionally, the giant resonance mode. 
In this mode each novel idea, once it appears, spreads  in an explosive manner in the 
theoretical community, sucking into itself a majority of active theorists, especially young 
theorists. Naturally, alternative lines of thought by and large dry out. Then, before this given idea bears fruit 
in the understanding of natural phenomena (due to the lack of 
experimental data and the fact that on the theory side 
crucial difficult problems are left behind, unsolved), a new novel idea arrives, the old 
one is abandoned, and a new majority jumps onto the new train. Note that I do not say 
here whether this is good or bad. This is just the fact of life of the 
present-day theoretical community which is largely deprived of reference points provided by experiment.

That's why such high expectations were associated with LHC. 

\section{Standard model}

Currently there are no direct experimental data contradicting SM. From
the discovery of neutral currents in 1973, through the precision electroweak measurements at LEP,  
to the discovery of the Higgs boson in 2012 -- everything we know today triumphantly confirms this model. Massive neutrinos and their mixing, which was absent in the earlier version, is naturally accommodated by SM. 
To this end no ``new physics" has been necessary. Note that  existence of axions {\em per se}, if 
confirmed in the future, will  require no restructuring of the model either, since the axions are neutral 
with regards to gauge interactions. One can just add them to the model as is.

The structure of the standard model is rather elegant and it is definitely self-contained. 
For all practical purposes we know  how our world operates at distances of the order 
of 10$^{-17}$ cm or larger (in many instances, to a high precision). 
However, the
curious mind never stops. Conceptual issues exist whose solutions lie beyond SM. 
First and foremost, the mass hierarchy.\footnote{One can also add an associated 
question of enormous suppression of the cosmological constant $\Lambda$.} 

There are rather many free parameters in the standard model, most of which are 
masses and quark (neutrino) mixing angles, which are clearly associated with the mass 
matrix. A complete lack of any semblance of universality in this sector is striking. Masses span the interval from $\sim 10^{-2}$ eV (for neutrinos) to $\sim 200$ GeV for $t$ quarks -- thirteen orders of magnitude. This is not the end of the story, however. Indeed, it is widely believed that the only natural scale in physics is set by the Planck mass $M_P$ which determines the strength 
(or should I say, weakness?)
of gravity interactions at low energies and the energy scale at which gravity becomes strong,  $M_P\sim 10^{19}$ GeV. This is, of course, assuming that no dramatic change occurs in physics going forward
 from the present-day $\sim 1$ TeV up to $M_P$, i.e. sixteen orders of magnitude. 

This idea -- that $M_P$ is the only genuine scale in physics --  is rather deeply rooted in the community, although
sometimes people still try to pose a question: ``What if this is not the case?"\footnote{ And rightly so. I remember that in the very beginning of my physics career it was not unusual to assume that the Fermi four-fermion interaction extends all the way up to its unitary limit, $E\sim G_F^{-1/2}$, and that $G_F^{-1/2}$ is the only scale relevant to weak interactions. Some theorists invested their efforts in exploration of this scenario. Needless to say, it was abandoned with the advent of the standard model with its $W$ and $Z$ bosons. Now we know that electroweak scale is set by
$M_W$, and at $E\sim M_W$ all cross sections are stabilized well below the unitary limit.}

If the natural scale of all masses is indeed set by $M_P$, then the hierarchy problem becomes awful. Not only 
are the masses in the matter sector  scattered over thirteen orders of magnitude, they are extremely small in the scale of $M_P$. 

Even if we accept for the time being that our understanding of physics is not 
ripe enough to explain the mass hierarchy, the next question to ask is whether or not this hierarchy is stable.
In  other words, if we set the mass parameters (measured in a certain well-defined way) more or less as they are in some approximation, will quantum corrections dramatically shift them from the initial values dragging them toward the Planck scale?

For fermions (quarks and leptons) the stability situation is not bad, provided that we are at weak coupling. Indeed, quantum corrections to masses are logarithmic and proportional to the original mass. Therefore, the expansion parameter 
$$
\frac{\alpha}{4\pi} \left( \log M_{\rm uv}/\mu_{\rm ir}\right) \ll 1
$$
 even in the worst case scenario in  which the ultraviolet cut-off parameter $M_{\rm uv}\sim M_P$.
 
 This is not the case, however, for the Higgs mass, or, alternatively, for the Higgs vacuum expectation value.
 Corresponding quantum corrections are quadratically divergent and, therefore, apparently drag these parameters toward the Planck scale, if there is no natural cut-off at a much lower scale. (Later I will say more about ``naturalness.")
 
 This is a bad situation. There are two ways out: new physics at a scale much lower than $M_P$, but not lower than 1 TeV or so (because below 1 TeV we see absolutely no indications of new physics), or extreme fine-tuning.
 The latter would mean that although each successive quantum correction produces a huge shift in mass, the shifts cancel in the total sum to a very high accuracy, so that the resulting overall shift is small or absent. Such a scenario is usually called ``unnatural." The criterion of naturalness is aesthetic, or, if you wish, philosophic. If you do not like it you can ignore it. Most people like it.
 
 Numerical situation with the cosmological constant $\Lambda$ might seem even worse. If $M_H/M_P \sim 10^{-17}$, for the cosmological constant we have $\Lambda^{1/4}/M_P \sim 10^{-31}$. In my opinion, there is no conceptual difference in  fine-tuning at the level of 17 or 31 orders of magnitude. One and the same, a hitherto unknown mechanism could be responsible in both cases.
 
 \section{Supersymmetry}
 
 Supersymmetry as a theoretical construction is known since early 1970s. Attempts at developing supersymmetry-based phenomenology started shortly after theoretical discovery of supersymmetry. In 1982 Witten pointed out that supersymmetry stabilizes the hierarchy problem. It introduces a new scale -- that of supersymmetry breaking $M_{/\!\!\!S}$ -- which, if low enough,  allows one to stabilize the Higgs boson mass. In supersymmetry, the quadratically divergent integral is   cut 
 off not by $M_P$ but by $M_{/\!\!\!S}$. In addition, the degree of fine-tuning in the cosmological constant
 is  reduced, since, as it became clear very early on, the cosmological constant vanishes in the limit of exact supersymmetry.
 
 Another reason for the advent of supersymmetry in phenomenology was the hope that it could provide us with 
 a sensible candidate for dark matter. If the $R$ parity is conserved, the lightest superpartner must be stable, and if the lightest superpartner is neutralino, the dark matter problem ($\sim 25$\% of the Universe's mass) could be solved by neutralinos.
 
 After Witten's publication, explorations in the framework MSSM, which became a basis for supersymmetric phenomenology,
  expanded in an explosive way. Although theoretically supersymmetry is a beautiful concept, the corresponding phenomenology was and still is less than elegant. Supersymmetry, if it exists, is definitely broken in nature. This breaking is parametrized by many free parameters. If in the standard model the number of free parameters is close to 20, in 
the  supersymmetric model it exceeds 100. Moreover, there are no deep theoretical reasons for the  $R$ parity conservation. If we allow $R$ parity   to be broken, extra free parameters appear and the dark matter motivation disappears, since in the absence of $R$ parity there are no stable superpartners. Worse than that, in the absence of $R$ parity we lose proton stability, generally speaking.
  
For many years supersymmetry-based  phenomenology was within the focus of theoretical research. By and large people closed their eyes to the above aesthetic drawbacks in the race for natural stabilization of the mass hierarchy. 

Now the discovery of the 125 GeV Higgs boson, and nothing else at LHC,  caught MSSM phenomenologists by surprise dramatically changing the overall picture and the state of minds within the community. A simple and elegant idea of a single scale $M_{/\!\!\!S}$ close to the electroweak scale turned out to be in contradiction with data! One could feel the mood of perplexity  in the audience. 

At the Lagrangian level MSSM predicts that $M_H< M_Z$, where $M_Z \approx 90\,$GeV is the $Z$ boson mass.
To elevate the Higgs mass to the level of 125 GeV one needs a very large radiative correction (not much smaller than the tree-level term). A natural solution is to make the stop mass (i.e. the mass of the $t$ quark superpartner) very heavy, perhaps from a few TeV to 10 TeV or heavier. 
In conjunction with the fact that superpartners are not seen at LHC, one must admit that (i) the scale of supersymmetry breaking is non-universal; (ii) it is likely to be very high, much higher than was expected 10 or even 5 years ago.
If  superpartners are much heavier than the electroweak scale 
then we are back to square one as far as the
  original problem of the hierarchy stabilization is concerned.  Already today we face the necessity of fine-tuning at the level of $10^{-2}$ or even $10^{-3}$. 

  Of course, people do not easily give up their dreams. They hasten to modify MSSM in a contrived  way to keep it viable.
  Split supersymmetry and  spread supersymmetry are just a few alternatives that (all of a sudden) regained popularity. 
 The version of MSSM which now goes under the name ``natural" (not to be confused with the original naturalness of the 1980s) is as follows: the first and second generation superpartners are assumed to be very heavy, so that there are no observable consequences from their existence whatsoever. The stop mass is fine-tuned to obtain the
 correct Higgs mass\footnote{The expected stop mass $m_{\tilde t} $ can be somewhat lowered, down to a few TeV,  at the price
 of introducing large and fine-tined $A$ terms.}
 ($m_{\tilde t} \sim 10\,$TeV). Then superpartners are not expected to be observed at LHC. Their appearance is deferred until an era of mythical ILC or some next-generation accelerator which may or may not materialize, certainly not soon 
 considering the present-day political climate. The original impetus for low-energy supersymmetry is 
 thus declared dead by the majority.\footnote{Some people still try to keep it afloat by developing contrived baroque-like aesthetically unappealing  modifications.}
 
 Is the current situation concerning phenomenological supersymmetry good or bad? I think it is good. The vicious circle of constrained MSSM has been broken. It is time to stop blindly scanning the parameter space and start thinking and developing new ideas. It is a great time for ingenious young researchers. I can compare it with two early years of my career, just before the advent of the standard model.
 
  \section{Theoretical supersymmetry}
 
This is an example of a complete success story. I use the word `theoretical' 
to differentiate from `phenomenological' supersymmetry  discussed 
above which, as I tried to convey, at the moment has a rather murky status.   
Theoretical supersymmetry proved to be a powerful tool with which to deal with quantum field theory, especially at strong coupling, a regime which was considered intractable for decades (with some exceptions in two dimensions).
Progress in this line of research, although slow,  is absolutely steady.

It was noted in the early 1980s that special holomorphy properties of supersymmetric gauge theories allow one to obtain exact results in the so-called protected sectors. The gluino condensate was calculated and the exact $\beta$ function was derived in this way. ${\mathcal N}=2$ supersymmetry turned out to be even more powerful in this respect. Continuous advances in this direction resulted in a revolutionary breakthrough in 1994, 
when the Seiberg-Witten solution
of ${\mathcal N}=2$ super-Yang-Mills was found. This was the first ever {\em analytical} demonstration of the dual Meissner effect as an underlying mechanism for quark (color) confinement. 

Equally important was the discovery of Seiberg's duality. In fact, it was first detected in supersymmetric QCD and then elevated to string theories, of which I will say a few words later. Of course, people knew from the early days of field theory that gauge symmetry is not a symmetry in the conventional meaning of this word, but rather is a redundancy in the theoretical description. Seiberg's duality explicitly demonstrated that different gauge theories, with distinct gauge groups, can lead to one and the same physics in the infrared. Needless to say, this can only happen if 
at least one theory in the dual pair is at strong coupling, so that the fields in the Lagrangian do not represent the asymptotic states of the theory. The Seiberg duality and its combination  with the Seiberg-Witten solution
 led to far-reaching consequences in the understanding of gauge theories.
 
 Another stimulating and promising development in theoretical supersymmetry is associated with extended objects such as domain walls (branes) and strings. 
 While studying dynamics of the BPS-saturated strings, 
 people came across a few surprises. First, dynamics on the string world sheet can be highly nontrivial. 
 Being strongly coupled, effective 
 world-sheet models are solvable (at least, some of them) because they are two-dimensional. And the solution of these two-dimensional models -- the most remarkable feature --  provides us with unambiguous (exact) information on
 aspects of  four-dimensional bulk theories. This phenomenon is now known as $2D-4D$ correspondence.
 
In certain instances supersymmetry-based results in conjunction with a $1/N$ expansion lead to exact predictions  for non-supersymmetric theories. Planar equivalence between ${\mathcal N}=1$ super-Yang-Mills and the the so-called orientifold theory presents the most clear-cut example. At $N=3$ the orientifold theory corresponding to 
${\mathcal N}=1$ super-Yang-Mills is just one-flavor QCD.

\section{Quantum chromodynamics}

It is now  hard to believe that before 1972 it was not uncommon to interpret field theory 
in  four dimensions as a set of  a few ``sacred" presumptions and prescriptions such as uniqueness of the vacuum state, the absence of vacuum condensates (a special procedure of normal ordering in the Lagrangian 
and various relevant composite operators
was always implied to ensure this property), and so on. By and large, quantum field theory was reduced to a theory of small field oscillations near zero, with the subsequent quantization of propagating waves and supplemented by a tricky and rather obscure procedure of renormalization.
It was little more than (if at all) a set of Feynman graphs.
 Because of the Landau zero charge no self-consistent field theories in four dimensions were known. Generally speaking, people (at least in  my surroundings) saw those who were stuck with field theory  as losers.\footnote{Arkady Vainshtein reminded me of important exceptions -- effective field theories, such as the theory of Goldstone bosons.}
 
Just for an illustration allow me to present a quotation from Andrey Linde's memoir: ``...The difference between weak and electromagnetic interactions arises after a non-zero vacuum average  $\langle\phi\rangle$ appears in the scalar field.  But according to quantum field theory, these averages should always exactly equal zero.  Many people at the time said the average $\langle\phi\rangle$ made no sense and that the spontaneous symmetry breaking mechanism  should simply be understood as a heuristic trick, necessary only to guess 
special relations
between masses and coupling constants 
of various fields
for which the theory becomes renormalizable." 

The advent of QCD changed all that. We learned that what you see in the Lagrangian is not necessarily what can be detected; that the vacuum structure can be complex, vacuum need not be unique; that small harmonic oscillations near vacuum are insufficient to explain strong dynamics; that nonperturbative physics is rich and important; that there is a variety of diverse regimes (or phases) that can be implemented in filed theory, such as Coulomb, Higgs, confinement, oblique confinement, conformal  and more; that the dual Meissner effect presents a typical mechanism leading
to  confinement of quarks (color); that Wilson's operator product expansion (OPE) can be adjusted to perfectly fit QCD, and then renormalization is readily (and trivially) understood as a process  of  evolution  from short to large distances.
OPE-based methods proved to be useful in many problems of practical interest, in particular, in heavy quark physics. 

In non-Abelian gauge theories one of the most profound and fruitful discoveries that shook the 
HEP community  was that of 't Hooft, who pointed out that 
$1/N$ is a (hidden) expansion parameter in QCD and Yang-Mills theories in general, corresponding to the expansion in topologies of the underlying Feynman graphs.  Thus, there emerged a natural -- albeit qualitative  -- correspondence between 
QCD and a string-like picture, with $g_s \sim 1/N$ where $g_s$ is the string coupling constant. Moreover, 
the domain wall tension in super-Yang-Mills was shown to scale as $N\sim 1/g_s$, which served as a basis for identification of these domain walls with the string theory branes.

A few times in the last two decades many believed, with excitement and enthusiasm,
 that the existing theory was at the verge of, if not the exact solution of QCD, at the very least its solution in the planar limit (i.e. $N\to\infty$ with the fixed 't Hooft coupling). I vividly remember these days. Alas ... these high expectations never came true. The range of natural phenomena that are described by QCD is so diverse and complex that such a universal solution seems  (to me) unlikely. Nevertheless, our understanding continues to grow in a non-revolutionary manner, in particular due to penetration of supersymmetry-based methods and proliferation of   $1/N$ expansions.
 
 \section{Grand unification}

If we let three gauge coupling, as they are known at our energies, run, assuming low-energy supersymmetry and nothing else at higher energies,  all three become equal to each other to a reasonable degree of accuracy at the energy scale $\sim 10^{16}\,$GeV. This scenario, which is also known as a Great Desert scenario, culminating in a beautiful unified 
$O(10)$ gauge group at the ``right" end of the desert, instead of $SU(3)\times SU(2)\times U(1)$ at our ``left" end, became deeply rooted in the minds of theorists during the three decades of its existence. 
The presence of a noncompact $U(1)$ is indeed an unpleasant theoretical feature of SM, for a number of reasons. 

Needless to say, in the absence of the low-energy supersymmetry the Great Desert scenario will have to be reconsidered. Will unification of the gauge couplings survive in some form?

\section{Large extra dimensions}

Ten years ago this was the hit of the day. String theory tells us that the number of dimensions may, generally speaking, be larger than four, for instance ten in the superstring theory. Then six extra dimensions must be compact. Such a solution (the Calabi-Yau compactification) was suggested 
by Candelas, Horowitz, Strominger, and Witten in 1984. This paper,  tacitly assuming that the size of the compactified dimensions is of the order of $M_P^{-1}$, initiated the attempts to get a realistic string theory 
in four dimensions.\footnote{ 
This pioneering work  was part of
the 1984 superstring revolution initiated by the discovery of the anomaly cancellation in type I 
string theory in ten dimensions by  Green and Schwarz.}

{\em A priori} this does not have to be the case: the patterns of compactification could be of a 
multistep/multiscale type, 
and quite contrived. It was suggested that at least one   extra dimension could have (in a sense) ``a macroscopic" size, with all matter fields trapped on a surface of a 3-brane with thickness of the order of $1/\,$TeV. 

The conceptual design was very attractive, transforming a gigantic hierarchy of the mass spectrum into a relatively modest hierarchy of a geometrical nature (the interbrane distances and 
brane thicknesses). Various interesting geometric separation mechanisms explaining the mass hierarchies, the pattern of the CKM matrix, suppression of unwanted flavor-changing decays and so on were worked out -- e.g.   fat
branes and  warped scenarios, to name just a few.

 As it often happens, however, the devil was in the details. Unfortunately no extra-dimension model which would be theoretically elegant and  concise on the one hand, and fully consistent with the existing phenomenology on the other hand, has ever appeared. Of course, the criterion `elegant' is relative, and one can argue whether a certain 
 model on the theoretical scene is elegant enough. What is unquestionable is the fact that now, ten years later, there are no indications of any of the large extra dimension models: nothing pointing  in this direction has come 
from LHC so far. Original enthusiasm seems to have faded away. Will it be resurrected? Only if  future LHC data will give it a chance. 
 
 The role of supersymmetry in the large extra dimension models is subsidiary, if at all; the Great Desert is gone, and unification of all three gauge couplings may or may not occur. In no way it can be considered as a {\em fait accompli}, 
 as is the case in low-energy supersymmetry. 

\section{String theory}

String theory appeared as an extension of the dual resonance model of hadrons in the early 1970, and by mid-1980 it raised expectations for the advent of ``the theory of everything" to Olympic heights. Now we see that these heights are unsustainable. Perhaps this was the greatest mistake of the string-theory practitioners. They cornered themselves by promising to give answers to each and every question that arises in the realm of fundamental physics, including the hierarchy problem, the incredible smallness of the cosmological constant, and the diversity of the mixing angles. I think by now the ``theory-of-everything-doers" are in disarray, and a less  formal branch of string theory is in crisis.\footnote{A more formal branch evolved to become a part of mathematics or  (in certain occasions) mathematical physics.}

At the same time, leaving aside the extreme and unsupported hype of the previous decades, we should say that string theory, as a qualitative extension of field theory, exhibits a very rich mathematical structure and provides us 
with a new, and in a sense superior,  understanding of mathematical physics and quantum field theory. 
It would be a shame not to explore this structure. And, sure enough, it was explored by serious string theorists. 

The lessons we learned are quite illuminating. First and foremost we learned that physics does not end in four dimensions: in certain instances it is advantageous to look at four dimensional physics from a higher-dimensional perspective. 
Surprisingly, a relatively simple geometric structure designed in higher dimensions -- let us call it string and brane engineering -- leads to highly nontrivial insights regarding the general (and sometimes even quite specific) aspects of supersymmetric gauge theories, and even two-dimensional sigma models. A significant number of advances in field theory, including miracles in ${\mathcal N}=4$ super-Yang-Mills, that we have
witnessed in the last decade or so came from the string-theory side. It turns out that a simple action -- abandoning the strive  to explain all of the world at the fundamental level, all at once, liberates string theory from the dead end it put itself in and places it onto a comfortable highway.

Much excitement was caused by the gauge-string duality, sometimes referred to as a holographic description. 
Since the 1980s Polyakov was insisting that QCD had to be reducible to a string theory in 4+1 dimensions. He followed this road step by step, for years practically alone, eventually arriving at  the conclusion that confinement in QCD could be described as a problem in quantum gravity. This paradigm culminated in 
 Maldacena's observation (in the late 1990's) that dynamics 
of ${\mathcal N}=4$ super-Yang-Mills in four dimensions (viewed as a boundary of a multidimensional bulk) at large $N$ can be read off from the solution of a string theory in the bulk. In particular, in the limit of the large 't Hooft coupling, the bulk string theory degenerates and becomes supergravity, which must be solved in the classical approximation.
Needless to say, searches for a classical solution of 
the supergravity equations are infinitely simpler than solving quantum field theory in the strong coupling regime.

Unfortunately (a usual story  when  fashion permeates physics), people in search of quick and easy paths 
to Olympus tend to overdo themselves. For instance, much effort is being invested in holographic description in condensed matter dynamics (at strong coupling). People pick up a supergravity solution in higher dimensions and try to find out whether or not it corresponds to any sensible physical problem which may or may not arise in a condensed matter system. To my mind, this strategy, known as the ``solution in search of a problem" is again a dead end. Attempts to replace    deep insights into relevant dynamics
with guesses very rarely lead to success. 

\section{Landscape} 

The idea of a landscape of vacua (which came from string theory) is probably the most dramatic change of paradigms  from the Newton times. In a sense, it was born out of desperation. The searches for a unique solution for our word, a unique string vacuum, ended in failure. No guiding principle was found to limit the number of vacua, let alone
to reduce the result to a unique vacuum which would contain in it all information currently available, including all mass hierarchies, all coupling constants, and so on. Because string theory proclaimed itself to be the {\em ultimate} theory, nothing short of that was acceptable. 

At this point  a U-turn occurred when some theorists suggested conversion of a failure of the
original program into a triumph. Observing that quasistable vacua of string theory, as they are  now known, are extremely abundant, 
they said: ``The more numerous they are the better!" If the number of vacua is $10^{500}$ (you can put in the exponent 1000, 10000 or any other large number you like) then various critical parameters  (all masses, coupling constants, generation numbers, chirality structure, gauge groups, etc.), being randomly scattered in these vacua, will give rise to a huge variety of distinct universes. Some of them (perhaps, just one or two) are quite exceptional. These are 
highly non-generic vacua  where we find our
worlds' parameters. We simply happen to live in such a vacuum. In other universes, life is impossible
and there are no theorists to derive laws of nature. 

Thus, there is no point in trying to understand the world order: the mass hierarchies, the smallness of the cosmological constant, 
the absence of the fourth generation, you name it. Nor will such attempts  be meaningful in the future.
It is all simply an environmental coincidence. Just take it as is and live happily ever after.

This is nothing other than  the anthropic principle in its extreme realization, with a religious (or philosophical, if you put it mildly) flavor. 

Indeed, even if this is true, we will never know. All ``extra" universes are causally disconnected from our own,
 so there is no physical way to confirm their existence or non-existence in experiment. So this part of the landscape paradigm is the act of belief in today's string theory, not supported by any evidence, and not to be supported by evidence in the future.

The second conclusion -- that one should abandon the search for a rational (non-environmental) understanding of the mass hierarchy/stability and the smallness of the cosmological constant -- may well be true
in a limited sense  but for a totally different reason than the landscape paradigm. It may well be that at the moment we do not have enough data to develop a theory.  Or, perhaps, a message has already been  sent to us but we failed to decipher  it. In other words,  the theory is not ripe enough
to offer an  explanation, in which case there is some hope that if the search is resumed in the future it will be more successful. 

\section{Curiosity}

Surprisingly, or perhaps not so surprisingly, the $H\to \gamma\gamma$ decay was discussed in quite a few talks.
A factor of 1.8 excess from the SM prediction (which -- I hasten to add -- is well within 2$\sigma$) caused a stir and gave rise to some speculations that it opened a window to new physics. Such a speculation seems premature. If only the discrepancy were at the level of 3$\sigma$ or more! Could VVZ and I have imagined, 33 years ago when we calculated this decay rate
and established its connection with the $\beta$ function, that this calculation would become an important reference point, a new physics counter of sorts?

\section{A lost generation?}

It is easy to estimate the total number of active high-energy theorists. Every day hep-th and hep-ph bring us about thirty new papers. Assuming that on average an active theorist publishes 3-4 papers per year, we get 2500 to 3000 theorists. The majority of them are young theorists in their thirties or early forties. During their careers many of them never worked on any issues beyond supersymmetry-based phenomenology or string theory. Given the crises 
(or, at least, huge question marks) we currently face
in these two areas, there seems to be a serious problem in the community. Usually such times of uncertainty as to the direction of future research offer wide opportunities to young people in the prime of their careers. It appears that in order to
take advantage of these opportunities a certain amount of reorientation and reeducation is needed. Will this happen?
  
\section*{Acknowledgments} 

I am grateful to Adi Armoni, Sasha Gorsky, Sasha Polyakov, and Arkady Vainshtein for useful comments.
This work  was supported in part by DOE
grant DE-FG02-94ER40823.

\end{document}